\documentclass[letterpaper,twocolumn,10pt]{article}
\usepackage{usenix2019_v3}

\usepackage{tikz}
\usepackage{amsmath}
\usepackage{amssymb}

\usepackage{tikz}
\usepackage{listings}
\usepackage{subfiles}
\usepackage{import}
\usepackage{transparent}
\usepackage{makecell}

\usepackage{subcaption}
\usepackage{enumitem}
\usepackage{multirow}
\usepackage{changes}

\newcommand{\name}{KeyMemRT}

\newcommand{\memImprAntace}{1.74x}
\newcommand{\timeImprAntace}{1.20x}
\newcommand{\memImprFhelipe}{1.16x}
\newcommand{\timeImprFhelipe}{1.73x}

\begin{document}
%-------------------------------------------------------------------------------

%don't want date printed
\date{}

\title{\Large \bf KeyMemRT Compiler and Runtime: Unlocking Memory-Scalable FHE}

%for single author (just remove % characters)
\author{
% Blind
Eymen Ünay\\
University of Edinburgh
% copy the following lines to add more authors
\and
Björn Franke\\
University of Edinburgh
\and
Jackson Woodruff\\
University of Edinburgh
} % end author

% \author{Eymen Ünay}
% \affiliation{%
%   \institution{University of Edinburgh}
%   \city{Edinburgh}
%   \country{UK}}
% \email{M.E.Unay@sms.ed.ac.uk}
%
% \author{Björn Franke}
% \affiliation{%
%   \institution{University of Edinburgh}
%   \city{Edinburgh}
%   \country{UK}}
% \email{B.Franke@ed.ac.uk}
%
% \author{Jackson Woodruff}
% \affiliation{%
%  \institution{University of Edinburgh}
%   \city{Edinburgh}
%   \country{UK}}
% \email{Jackson.Woodruff@ed.ac.uk}

\maketitle

% Your abstract text goes here. Just a few facts. Whet our appetites. Not more than 200 words, if possible, and preferably closer to 150.
\begin{abstract} 
  Fully Homomorphic Encryption (FHE) enables privacy preserving computation but it suffers from high latency and memory consumption.
  The computations are secured with special keys called rotation keys which often take up
  the majority of memory. 
  In complex FHE applications, these rotation keys can cause a large memory bottleneck limiting program
  throughput.
  Existing compilers make little effort to solve this problem, instead relying on systems
  with massive memory availability.
  This resource requirement is a barrier
  to FHE uptake because optimizing FHE programs by hand
  is challenging due to their scale, complexity and
  expertise required.

  In this work, we present KeyMemRT; an MLIR based compiler and runtime framework that \textit{individually} manages rotation key lifetimes to lower memory utilisation and to allow arbitrary number of rotation indices to be supported without memory bloating.
  KeyMemRT relies on dataflow analysis to determine
  key lifetimes and is the first FHE compiler to provide automatic
  key management, handle fine-grained key-mangement and
  manage boostrap keys.
  We implement frontends for Orion and HEIR 
  and show improvements over state-of-the-art FHE compilers. KeyMemRT achieves memory reduction of \memImprAntace{} and a speedup of \timeImprAntace{} over ANT-ACE, and memory reduction of \memImprFhelipe{} and a speedup of \timeImprFhelipe{} over memory-optimized compiler Fhelipe.
  We provide KeyMemRT as a post-optimizing compiler that can be targeted by any FHE compiler. 
\end{abstract}

\section{Introduction}
%problems
%memory problem
%Existing work challenges
%This work
%Contribs

\begin{figure}
	\includegraphics[width=\columnwidth]{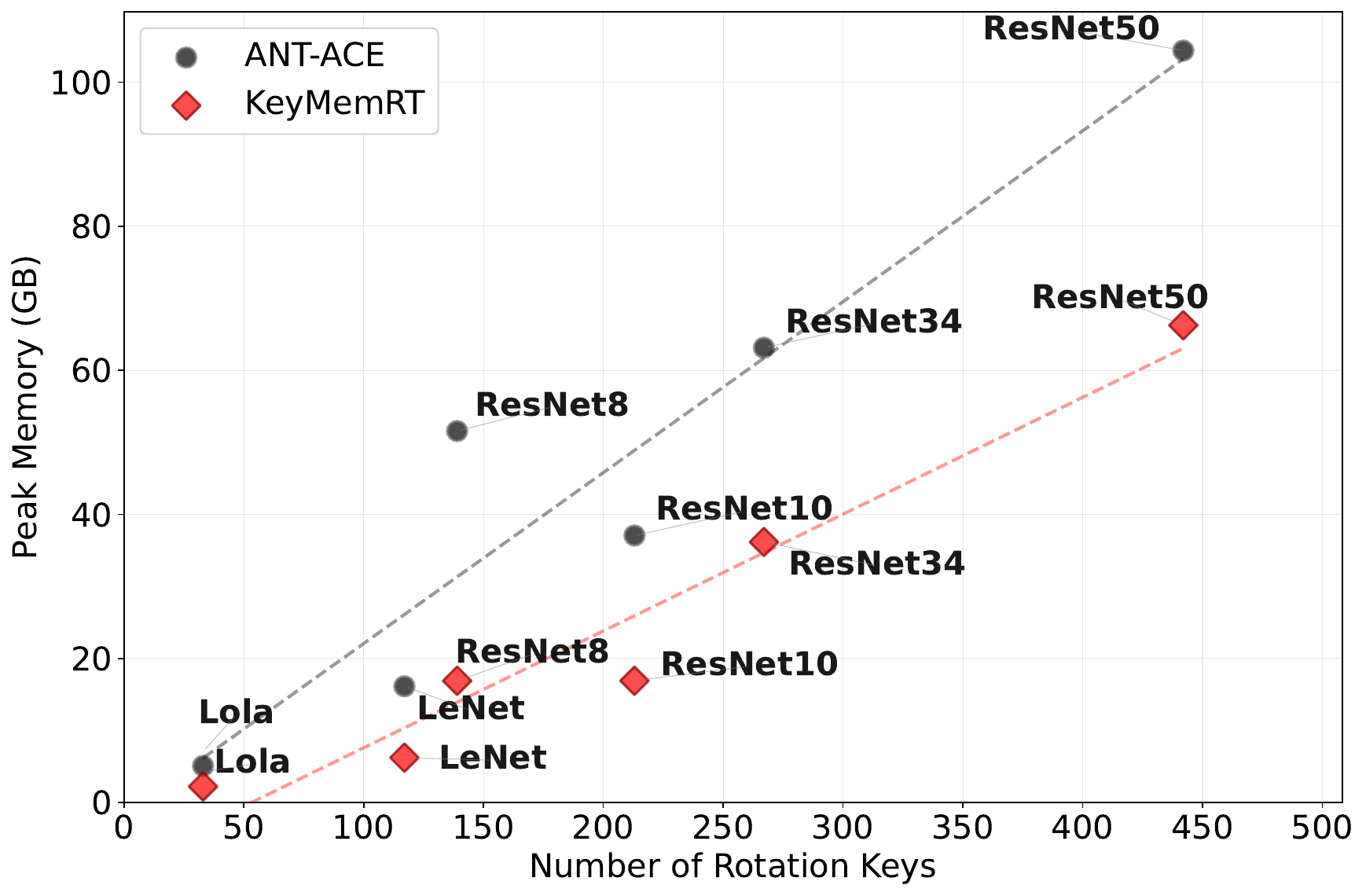}
   \caption{
Currently the memory requirements from larger numbers of keys is not scalable. KeyMemRT provides lower memory growth without overhead allowing FHE programs to scale better.}
   
	\label{fig:keys_size_scatter}
\end{figure}

%TODO show system impact like RAM disk 1GB space saved
\begin{figure*}
	% \includegraphics[width=\columnwidth]{fig/code_motivation_example_draw.pdf}
	% \begin{subfigure}{0.33\textwidth}
       % \def\svgwidth{\textwidth}
       % \import{fig/}{code_motivation_example_draw.pdf_tex}
	% \caption{}
   % \end{subfigure}
	   % the text positioning in the fig was looking weird to me
	   %\def\svgwidth{\textwidth}
	   %\import{fig/}{memory_motivation_example.pdf_tex}
	\includegraphics[width=\textwidth]{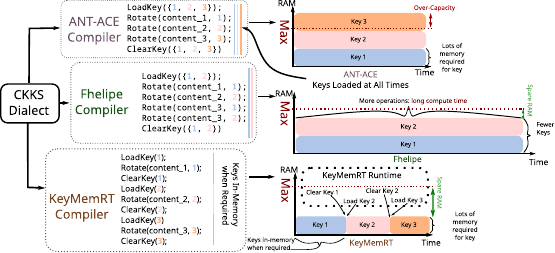}
   \caption{
	   The \name{} compiler takes the \texttt{ckks} dialect as input
	   and optimizes the code to keep keys out of memory as much
	   as possible by unoverlapping key livenesses.
	   The \name{} runtime manages keys, loading them in and
	   clearing them as required to keep memory utilization
	   down.
	   Fhelipe uses decomposition to reduce the number of keys
	   required, but this results in long runtimes. ANT-ACE
	   loads all the keys in, which results in fast computation
	   but wasted memory.
	   % \name{} performs a number of compute-time optimizations to
	   % overcome the overhead of loading and clearing keys.
   }
	\label{fig:motivation_example}
	\label{fig:memory_example}
\end{figure*}

The increasing trend in data collection and processing raised concerns over privacy in recent years. Current measures are not effective with the increasing number of data leaks and privacy violations. Fully Homomorphic Encryption (FHE) is the only technology bringing privacy into the computation level by keeping the data in encrypted state at all times. This allows sectors with sensitive data such as medical or financial to 
 make use of cloud computing without breaching privacy guarantees. 
 However, FHE faces a barrier to real-world uptake because of its
 excessive compute and memory requirements which are
 challenging to optimize by hand.

%Rotation operations are one of the three primitive operations
Rotation operations are one of the most expensive primitive operations
in batched FHE
schemes and are necessary for intra-vector computations and bootstrapping. However,
it is impossible to store all the rotation keys for every possible rotation index
as they can take 10s of TBs~\cite{leeRotationKeyReduction2023a}.  Figure \ref{fig:keys_size_scatter} shows that FHE programs fail to have memory-scalability with existing
FHE compilers as the memory increases rapidly with the number of keys. Na\"ive implementation of simple kernels such as matrix-vector
multiplication require that keys scale with the problem size, exacerbating this problem.
Existing compilers treat these keys as a single entity, which
creates a memory-scalability problem.
Managing these keys efficiently requires fine-grained control and large-scale program
transformations that existing FHE compilers fail to even attempt.
%JCW: I think we need a ref to fig 1 in this para somewhere.
%These challenges with key set size means that FHE suffers from high memory consumption
%and latency, which limits throughputs, even on large servers.

Existing strategies to unlock memory-scalable FHE
compromise on program speed.
%Orion~\cite{ebelOrionFullyHomomorphic2025} uses a hand-written mechanism that
%loads sets of keys for matrix multiplication (matmul) kernels on demand
%reducing average memory consumption but this is not portable to different FHE applications.
Fhelipe~\cite{krastevTensorCompilerAutomatic2024} generates fewer keys and chains rotations
as sums of power of two, reducing total memory
consumption, but slowing down the program considerably.
%Most FHE compilers do not attempt key management
Most FHE compilers target libraries which do not support
key management and keep rotation
keys in global maps which are kept in memory throughout the execution of
the program~\cite{badawiOpenFHEOpenSourceFully2022,mouchetLattigoMultipartyHomomorphic2020,chenSimpleEncryptedArithmetic2017}.
%where the server loads all the rotation keys at once.
%Some FHE compilers attempt to get around the memory problem caused by rotation keys. ANT-ACE generates the exact rotation keys that the program uses and does no key management.
Neither of these approaches scales well to large FHE
programs, where minimising memory consumption
requires fine-grained key management. In fact,
existing compilers lack sufficiently expressive IRs
and sufficiently detailed analyses to tackle the
fine-grained key-management problem.

In this work, we introduce \textit{KeyMemRT}, an FHE compiler and runtime framework that manages rotation
keys individually.
We introduce the \texttt{kmrt} MLIR dialect, and use this dialect to
represent key liveness and key movement hints ~\cite{lattnerMLIRScalingCompiler2021}.
Our compiler treats rotation keys as dataflow elements and
is the first to perform fine-grained, automated analysis 
on large FHE programs.  It inserts key movement
hints and supports state-of-the-art key-reduction techniques.
The runtime uses these compiler-inserted hints to keep keys in memory when they are needed, reducing the
memory requirements  by \memImprAntace{} over ANT-ACE and reducing the latency by \timeImprFhelipe{} over Fhelipe in a server environment.

In summary, we contribute:
\begin{itemize}[nolistsep,noitemsep]
	\item A new MLIR dialect called \texttt{kmrt} with a new type for rotation keys
		and memory management operations exposing keys as IR elements.
	\item An MLIR-based compiler built on \texttt{kmrt} dialect
		that tracks key liveness,
		optimizes bootstrap keys and reduces key numbers.
	\item A runtime that manages these keys, loading keys in when they are needed
		 with a prefetching mechanism and clearing them out.
	\item We package these into a single tool,
		\name{}, which automatically enables memory
		 reduction of \memImprAntace{} over ANT-ACE and latency reduction of \timeImprFhelipe{} over Fhelipe.
\end{itemize}

		% liveness analysis to track keys,
		% identifying when keys are needed in memory and
		% when they can be cleared from memory.

\subsection{Motivating Example}
%DATA 20% is Resnet opt in 90GB machine, 94% Alexnet Prefetch in the same machine
Rotation keys require vast regions of memory
in FHE programs, taking up 20-94\% of the memory utilization in 
our benchmarks and requiring 100s of GB\@.
This limits the scalability of FHE programs, because it presents
a fundamental limit to the vectorization of programs and to the compute platforms.
In \name{}, we use liveness analysis and prefetching to load
keys in on-demand. 
% Figure~\ref{motivation_example} shows an
% example of how existing approaches manage keys and how \name{}
% manages keys.

\subsubsection{What is Memory-Scalable FHE?}
FHE programs have excessive memory requirements,
often in the 100s of GBs.  To support these programs
effectively in the real-world, programs need to use the memory
available to them in the most efficient manner possible.

In this paper, we discuss memory-scalability as a property that
enables arbitrarily large FHE programs to run without additional memory burden --- enabling servers to run
larger FHE programs than currently possible.
Existing approaches are either not memory-scalable (ANT-ACE), or
introduce significant compute overhead (Fhelipe).

\subsubsection{ANT-ACE: No Key Management}
ANT-ACE takes an approach representative of most FHE compilers, where
keys are stored in-memory throughout program execution.
Figure~\ref{fig:motivation_example} shows this strategy at the top.
Here, we can see that all keys
are live throughout the program, even though they are not needed at the same time.
This causes memory consumption challenges that the majority of FHE compilers
face.
% Most FHE libraries store rotation keys in global maps which are kept throughout the execution of the
% program \cite{badawiOpenFHEOpenSourceFully2022, mouchetLattigoMultipartyHomomorphic2020, chenSimpleEncryptedArithmetic2017}. In a practical client-server interaction, client generates and sends all of the rotation keys and other
% cryptographic metadata required by the server.
% The server receiving the rotation keys in a file, deserializes it in the beginning of execution and can execute the rotations.
% This means all of the rotation keys are loaded into memory and kept in there until the program ends.

% TODO -- need to cover orion and fhelipe also here.

\subsubsection{Fhelipe: Binary Decomposition}
Fhelipe ~\cite{krastevTensorCompilerAutomatic2024} is a compiler focusing on data packing in
FHE tensor applications.
% It starts from a high level Python based description of the models
% lowering them into low level FHE operations and executing them in Lattigo calls.
Figure~\ref{fig:motivation_example} shows Fhelipe's approach to memory-saving in the middle:
it uses fewer keys and instead performs more compute operations.

Fhelipe employs no explicit key management. It generates keys for a pre-determined set
of indices for every program. The set includes indices of powers of two in positive and negative 
direction of rotations. The runtime then translates a rotation of arbitrary index into a chain
of rotations with indices that sum to that index.
While this resolves the memory challenges, it comes at a high price
in terms of compute time and this approach scales poorly with the number
of keys.

\subsubsection{\name{}: Fine-Grained Key Liveness Analysis}
\name{} tracks the liveness of rotation keys using
compiler analysis and overcome the overheads using compiler optimizations.
Figure~\ref{fig:motivation_example} shows how the KeyMemRT compiler
defines and splits live ranges of keys and the KeyMemRT runtime uses
the hints to load and clear keys as required.
This keeps the memory usage low, and the computation time
fast, enabling memory-scalability.

Table~\ref{tab:feature_comparison} shows how \name{}'s features
compare to the ANT-ACE and Fhelipe. \name{} is the first
compiler to do automatic key management, fine-grained key management
and boostrap key management. In addition, similar to
Fhelipe, \name{} supports rotation decomposition.

%As shown in Figure \ref{fig:motivation_example},
%the global liveness ranges for the rotation keys
%for index 1 and 2 are for the entire program even when they are not needed.
%KeyMemRT on the other hand assigns lifetimes which halves the number of keys needed to be stored in memory.
%%KeyMemRT makes use of the fact that the liveness range can be split and be trimmed depending on the program structure.
%Because KeyMemRT relies on compiler techniques, this liveness information
%can be determined for each program,
%and be used by the runtime to manage the keys
%and load them in as required.  This is shown in
%Figure~\ref{fig:memory_example}, where keys are loaded
%into memory from disk when live and cleared from memory when
%not in use.

% This insight enables the memory savings of 1.6x that
% we report in the results section.

\begin{table}[h]
  \centering
  \setlength{\tabcolsep}{2pt}
  \begin{tabular}{|l|ccc|c|}
    \hline
     & ANT-ACE & Fhelipe & \textbf{KeyMemRT} \\
    \hline
    \makecell[l]{Automatic\\Key Management} & $\times$  & $\times$ & $\checkmark$ \\
    \hline
    \makecell[l]{Decomposition\\Support} & $\times$ & $\checkmark$ & $\checkmark$ \\
    \hline
    \makecell[l]{Fine grained\\key management} & $\times$ & $\times$ & $\checkmark$ \\
    \hline
    \makecell[l]{Bootstrap\\key management} & $\times$ & $\times$ & $\checkmark$ \\
    \hline
  \end{tabular}
  \caption{State-of-the-art FHE compilers have partial support for dealing with rotation keys but none provide a fine-grained system. KeyMemRT is the first general-purpose automatic key management compiler.}
  \label{tab:feature_comparison}
\end{table}

%TODO more system perspective

%TODO new subsection here

% \newcommand{\baseMethod}{Global }
% \newcommand{\ourMethod}{KeyMemRT }

\section{Background}
Fully Homomorphic Encryption (FHE) performs computation over encrypted data which allows privacy preserving computation. While encryption and decryption require keys such as the public key and the private key, FHE introduces \textit{evaluation keys} which are necessary for computation involving encrypted data \cite{marcollaSurveyFullyHomomorphic2022}.
\begin{align}
&c_f = \text{Eval}_{\text{evk}}(f, (c_1, \ldots, c_t)) \\
&\text{Dec}_{\text{sk}}(c_f) = f(m_1, \ldots, m_t)
% \label{eq:homomorphic_property}
\end{align}

Batched FHE schemes such as BGV, BFV and CKKS work with vectors of slots with encrypted data ~\cite{brakerskiFullyHomomorphicEncryption, brakerskiFullyHomomorphicEncryption2012, fanSomewhatPracticalFully2012b, cheonHomomorphicEncryptionArithmetic2017}. However, FHE vectors do not allow extracting a specific slot and the only way to perform operations between different slots of a vector is to perform rotation. Rotation operations cyclically shift vector slots by a given index.
% Primitive operations of multiplication and rotation in batched schemes of FHE such as BGV, BFV and CKKS, require evaluation keys. However, rotation operations require a unique evaluation key for every index which makes up most of the evaluation keys. If the application requires a large amount of unique rotation indices, the memory overhead can be in the magnitude of hundreds of GBs. The memory overhead hinders practical adoption of FHE and can make it unusable in edge scenarios and accelerators with limited memory capacity.

\subsection{Rotation Keys}
%rotation keys -> key switching keys
% Relinearization keys
% ...more

FHE relies on several types of evaluation keys to enable
computation involving encrypted data \cite{marcollaSurveyFullyHomomorphic2022}. This is because when an operation 
such as multiplication or rotation is applied, it results in a ciphertext 
that is not decryptable by the original secret key. To recover the message
with the original secret key, key-switching operation is performed~\cite{cheonHomomorphicEncryptionArithmetic2017}. However,
key-switching operation requires keys for every specific transformation. 
Key-switching keys are generated from secret key and should be provided by
the data owner such as the client.
While operations such as multiplication or conjugation can use a single key,
rotation operations require a different key for every rotation index. 

FHE libraries strictly require a unique rotation key for a specific rotation index. If an FHE programmer wants to use a set of rotation indices in the program, they should make sure all of the keys for the indices are generated and available to the library.

Because key-switching is coupled to operations, rotation keys
are only present in FHE schemes where rotation operations are defined. Batched 
FHE schemes such as BGV, BFV and CKKS have all rotation as 
a primitive operation. Because of its high performance, \name{} focuses on 
the CKKS scheme which is also suitable for approximate applications such as Machine Learning (ML). The
key management of \name{} is based on the CKKS scheme operations and its memory 
reduction is applicable to all FHE libraries.

\subsection{Scalability of Rotation Keys}\label{Sec.Background.Scalability}
Rotation operations are necessary when different slots of the FHE vector need to 
interact in computation.
Each different slot (or index) into the vector requires
a unique key, which introduces a scalability challenge.
These inter-vector dependencies are highly algorithm specific but even
simple applications such as summing all slots of an FHE vector, require careful design 
to minimize the number of time- and memory-costly rotation operations.
In many cases, dependencies are inherent to the algorithm and cannot be removed.

%TODO simplify the sorting explanation
% \added{New algorithms are designed for FHE since conventional algorithms are not designed with constraints such as multiplicative depth or number of bootstraps.
% These algorithms push for performance by better effectively using the large number of available vector slots however this rises the number of slot interactions and the keys.}

For example, most high performing FHE sorting algorithms 
involve rotation indices ranging from $1$ to the input array size $N$ since every vector slot needs to be aligned and compared with each other requiring $N$ keys~\cite{mazzoneEfficientRankingOrder2025, kimOptimizedRankSort2025}. Similarly state-of-the-art FHE plaintext matrix-vector multiplication
algorithms such as diagonal order require rotation indices ranging from $1$ to the input 
matrix size $N$ and therefore $N$ keys~\cite{haleviFasterHomomorphicLinear2018}.
This is because a matrix mapping to 1D FHE vector requires every matrix slice to be in different slot positions than other slices. Due to the performance benefits
of FHE level vectorization, most FHE algorithms are designed to pack more data in a single FHE 
vector causing more slot interactions, rotation indices and thus more keys.

As FHE becomes more usable, larger inputs are used and more complex programs are run, the memory problem 
becomes more evident. It is not practical to use alternative algorithms instead due to the significant performance
benefit of efficient FHE vectorized algorithms. We pinpoint this issue as a roadblock to FHE memory scalability and in this
work provide an alternative way to preserve performance while lowering memory.

\subsection{Bootstrap Keys}
Bootstrap operations are composite operations that execute FHE programs including rotations
internally.
They refresh the computation budget of the FHE variables by increasing ciphertext levels and 
are essential to make an FHE program run indefinitely.
However, it is a time consuming and memory heavy operation which is why
libraries provide specially crafted high-performance kernels for it.

While libraries define bootstrap keys in APIs, they are mostly rotation keys and additional keys such as the conjugation key. As explained in Section \ref{Sec.Background.Scalability}, extending programs with  their own sets of rotation keys either keeps the number of keys same or increases. In the case of bootstraps, currently even if a program uses a bootstrap operation once it has to keep all of its keys in memory adding on to the memory burden.
 Despite the large number of keys, 
there has been no study of bootstrap key management to reduce memory consumption.

\subsection{Existing Memory Management Strategies}

FHE programmers, compilers and libraries have attempted to deal with the memory problem of rotation keys in various ways.

\subsubsection{Rotation Chaining}\label{Sec.Background.RC}

The existing most common approach to reduce the memory overhead of rotation keys is to chain base rotations to arrive at the rotation indices of the program. Since rotation operation is based on automorphism, rotations are distributive over addition and multiplication \cite{haleviFasterHomomorphicLinear2018}. A rotation $\text{Rot}(\text{c}, \theta)$ with target rotation index $\theta$ over input data $c$, can be decomposed into multiple rotations $\text{Rot}(\text{c}, \theta) = \text{Rot}(\text{Rot}(c, \theta_0), \theta_1)$ such that $\theta = \theta_0 + \theta_1$. Chaining arbitrary number of rotations is possible as long as the decomposed indices add up to the target index $\theta = \theta_0 + \theta_1 + \ldots + \theta_n$. A target index $\theta$ without a rotation key can be computed by chaining rotations with indices supported by rotation keys $\text{Rot}(\text{c}, \theta) = \text{Rot}_{evk_1}(\text{Rot}_{evk_0}(c, \theta_0), \theta_1)$.

\textbf{Baby-Step Giant-Step (BSGS)}:  Baby-Step/Giant-Step (BSGS)~\cite{haleviFasterHomomorphicLinear2018} decomposes a linear sequence of rotation indices into two rotations of giant and baby steps. If the target rotation indices are assumed to be a sequence $1\ldots N$, then the giant step size is defined to be $g = \lceil\sqrt{N}\rceil$. The giant step, divides the sequence into equal parts as $g, 2\times g, \ldots, \left\lceil N/g\right\rceil \times g$. For indices in the subsequences, baby steps are used which are defined as a sequence up to the giant step as $1, \ldots, g-1$. BSGS is proposed to be used in diagonal order matrix-vector multiplication which requires rotation indices $1\ldots N$.
BSGS results in around $2 \times \sqrt{N}$ number of keys instead of $N$. While BSGS guarantees that the maximum number of chained rotations for any index will be two, its memory overhead increases rapidly as the number of target rotation indices increase.

\textbf{Powers-of-Two Chaining}: This decomposition expresses any integer rotation index as a sum of powers of two. Every addend corresponds to a chained rotation while the rotation indices are always known to be powers of two beforehand. This approach fixes the memory consumption to a handful of integers but makes compromise on the speed of the program as rotation operations latency is effectively multiplied by the chain length.

\section{KeyMemRT: System Overview}

\begin{figure}
	\includegraphics[width=\columnwidth]{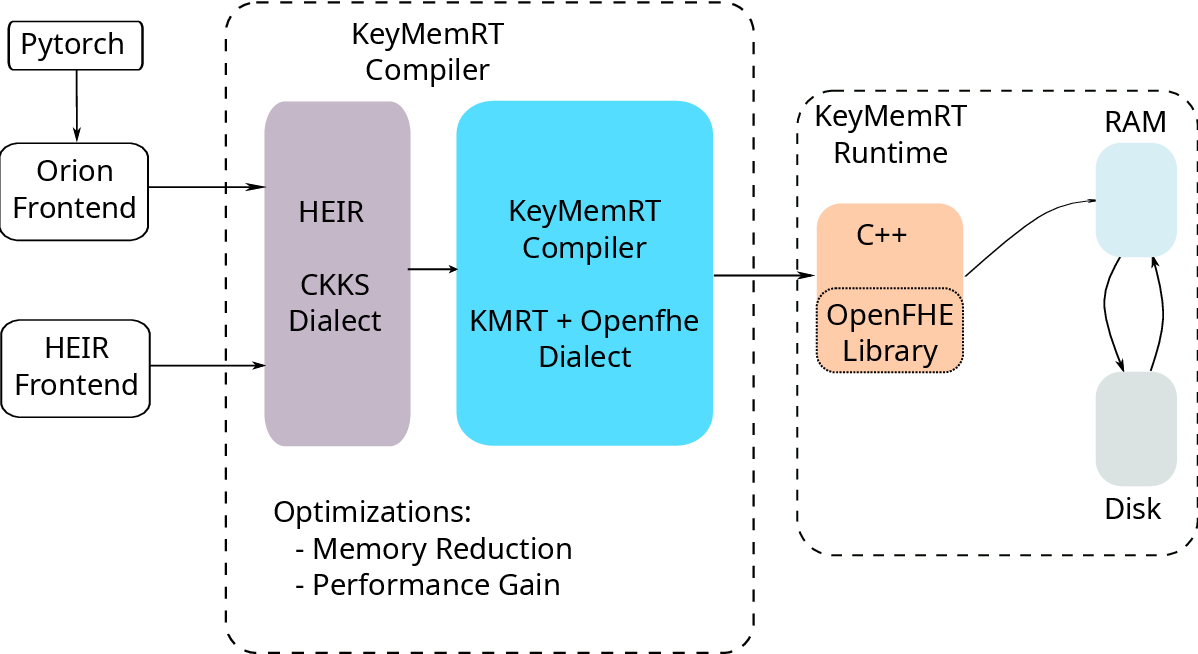}
	\caption{Design of KeyMemRT.  The MLIR-based compiler
		works with the \texttt{openfhe} dialect and inserts
		key-management operations in our \texttt{kmrt} dialect.
		The runtime uses these indications to manage keys.
	}
	\label{fig:keymemrt_diagram}
\end{figure}

KeyMemRT is a post-optimizing compiler which optimizes programs
lowered to FHE primitive operations.
Figure \ref{fig:keymemrt_diagram} shows the system level flow of KeyMemRT.
It takes the library-independent HEIR~\cite{aliHEIRUniversalCompiler2025} \texttt{ckks} dialect
as input and produces \texttt{openfhe} and \texttt{kmrt} dialects.
The final MLIR is then translated into its C++ equivalent 
and the runtime manages memory during execution.

We use the Orion frontend to lower Pytorch models to FHE primitive operations \cite{ebelOrionFullyHomomorphic2025, paszkePyTorchImperativeStyle2019}.
Our custom translator converts Orion's internal compiler IR into the HEIR
middle-end dialect \texttt{ckks}.
To enable translation from Orion to 
HEIR, we introduced high level ops in \texttt{ckks} such as \texttt{ckks.linear\_transform}
and \texttt{ckks.chebyshev}. We introduced our own lowerings for the new ops and used HEIR's existing lowerings for the rest of the \texttt{ckks} ops
to get \texttt{openfhe}
dialect.

The KeyMemRT compiler takes the output \texttt{openfhe} dialect of HEIR and
starts its compiler pipeline. It analyzes and transforms the program rotation operations
and inserts \texttt{kmrt} dialect typed values and operations. After the optimizations,
the output MLIR of KeyMemRT compiler is translated to C++ and compiled to executables.
The KeyMemRT runtime orchestrates key management during execution by loading
the keys from disk at the right time and clearing them out.

\section{KeyMemRT Compiler}
The compiler exposes rotation keys as intermediate representation (IR) elements, performs dataflow analysis over the program and tracks rotation keys. 
KeyMemRT compiler includes several lowering and optimization passes for reducing 
the number rotation key loads.

\begin{figure*}
      \def\svgwidth{0.99\textwidth}
       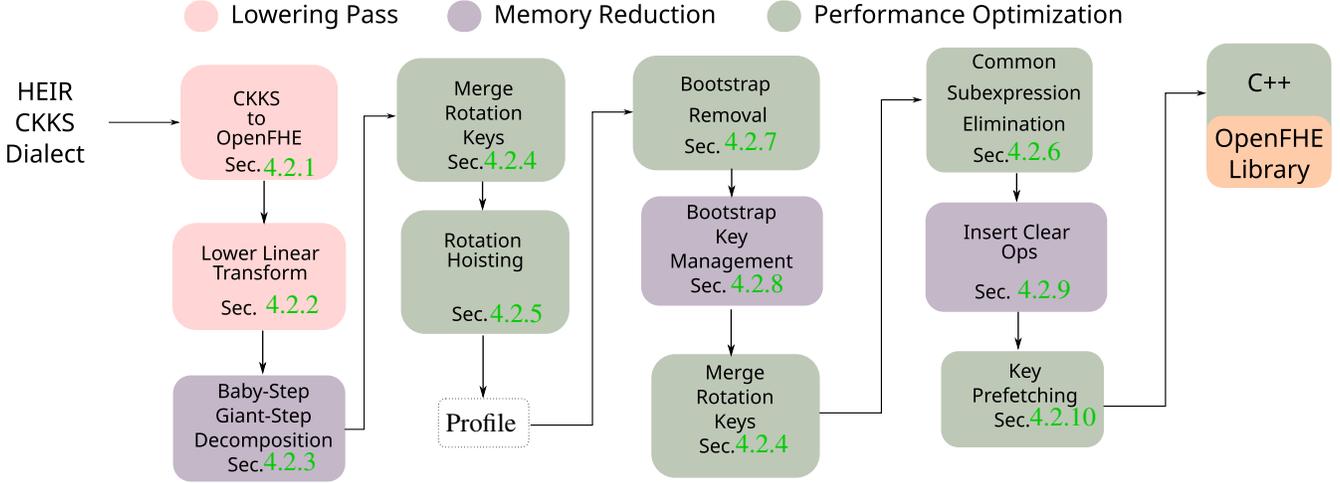
       \caption{The full pipeline of the KeyMemRT compiler. CKKS to OpenFHE (\ref{Sec.Comp.CtO}) and Lower Linear Transform (\ref{Sec.Comp.LLT}) passes perform lowering and generate the \texttt{kmrt} operations. Baby-Step Giant-Step (BSGS) Decomposition (\ref{Sec.Comp.BSGS}) reduces the number of keys in the program as explained in Section \ref{Sec.Background.RC}. First Merge Rotation Keys (\ref{Sec.Comp.MRK}) pass manages rotation keys of explicit rotations by the user and the decomposed rotations from BSGS. Rotation Hoisting (\ref{Sec.Comp.RH}) speeds up rotations applied to common inputs. After compiling and profiling at this stage, the FHE profile is used in Bootstrap Removal (\ref{Sec.Comp.BR}) to speed upthe program. Bootstrap Key Management (\ref{Sec.Comp.BKM}) materializes the keys used opaquely by the library to the IR level. Second Merge Rotation Keys (\ref{Sec.Comp.MRK}) pass merges key livenesses of the bootstrap related rotation keys with the explicit rotations. Common Subexpression Elimination (\ref{Sec.Comp.CSE}) removes cleartext dead code resulted from the \texttt{kmrt} optimizations. Insert Clear Ops (\ref{Sec.Comp.ICO}) reduces memory consumption by clearing plaintexts and ciphertexts after last use. Lastly, Key Prefetching (\ref{Sec.Comp.KP})  places hint operations that will inform the runtime. The final generated code is a C++ program with OpenFHE API and KeyMemRT runtime calls.
	   }
	\label{fig:compilerDiagram}
\end{figure*}

% Full pipeline example 
% $(MLIR_DIR)/%_openfhe_staticopt.mlir: $(MLIR_DIR)/%.mlir
% 	@echo "=== Static optimizations for $* ==="
% 	$(KEYMEMRT_OPT) \
% 		--ckks-to-lwe --lwe-to-openfhe \
% 		--lower-linear-transform \
% 		--symbolic-bsgs-decomposition --kmrt-merge-rotation-keys \
% 		--annotate-module="backend=openfhe scheme=ckks" \
% 		--openfhe-configure-crypto-context --openfhe-fast-rotation-precompute \
% 		$< > $@
% 	@echo "✅ Generated $@"
%
% $(MLIR_DIR)/%_openfhe_pgoopt.mlir: $(MLIR_DIR)/%_openfhe_staticopt.mlir $(PROFILE_DIR)/%_staticopt_profile.txt
% 	@echo "=== PGO1 optimizations for $* ==="
% 	$(KEYMEMRT_OPT) \
% 		--profile-annotator="profile-file=$(shell pwd)/$(PROFILE_DIR)/$*_staticopt_profile.txt" \
% 		--remove-unnecessary-bootstraps --bootstrap-rotation-analysis --kmrt-merge-rotation-keys \
% 		--cse --openfhe-insert-clear-ops  \
% 		$< > $@
% 	@echo "✅ Generated $@"

\subsection{MLIR based \texttt{kmrt} Dialect}
%TODO should we visualize this somehow  figure \ref{fig:lowering_rot} might be good
We propose a new MLIR dialect called \texttt{kmrt} which represents
key management directly.
%To represent key management as a compiler intermediate representation for analysis and
%transformations, we propose a new MLIR dialect called \texttt{kmrt}. 
It introduces new 
types and operations specific to key management and can be reused with any FHE IR supporting rotations.
These new operations are shown in Table~\ref{tab:kmrt_operations}.

\begin{table}[htbp]
\centering
\caption{MLIR \texttt{kmrt} Dialect Operations and Types}
\label{tab:kmrt_operations}
\begin{tabular}{|l|l|}
\hline
\textbf{Type} & \textbf{Parameter} \\
\hline
\texttt{!kmrt.rot\_key<index>} & \texttt{index}: integer \\
\hline
\hline
\textbf{Operations} & \textbf{Types} \\
\hline
\texttt{kmrt.prefetch\_key index} & \texttt{index}: integer \\
\hline
\texttt{kmrt.load\_key index} & \texttt{index}: integer \\
 & $\rightarrow$ \texttt{!kmrt.rot\_key} \\
\hline
\texttt{kmrt.clear\_key rk} & \texttt{rk}: \texttt{!kmrt.rot\_key} \\
\hline
\hline
\textbf{Helper Operations} & \textbf{Types} \\
\hline
\texttt{kmrt.use\_key rk} & \texttt{!kmrt.rot\_key} \\
 & $\rightarrow$ \texttt{!kmrt.rot\_key} \\
\hline
\texttt{kmrt.assume\_key index} & \texttt{index}: integer \\
 & $\rightarrow$ \texttt{!kmrt.rot\_key} \\
\hline
\end{tabular}
\end{table}

\textbf{Types}
\texttt{kmrt} provides the type \texttt{rot\_key} which is parameterized with the rotation index.
This way the compiler can identify what index a given key supports since keys generated
for different indices are not compatible in FHE\@. HEIR \texttt{openfhe} dialect's rotation operation is extended to require another operand typed with \texttt{rot\_key}. 

\textbf{Operations}
\texttt{kmrt} provides operations that handle \texttt{rot\_key} typed values. The
operation \texttt{kmrt.load\_key} gets the rotation index requested and,
returns the rotation key value. The operation \texttt{kmrt.clear\_key} takes the rotation key value as an argument. \texttt{kmrt.use\_key} and \texttt{kmrt.assume\_key} are IR specific operations to switch scope of a key. This flow allows every rotation operation
to have a live key associated with them. This type safety design helps in building local scopes, allows using dataflow analysis and prevents errors in optimizations.

\subsection{KeyMemRT Compiler Passes}
The KeyMemRT compiler lowers to, and optimizes on the \texttt{kmrt} dialect.
Lowering involves introducing linear transform operators and
introducing explicit rotation key management.  Optimizations follow,
which involve inserting key management strategies, and then
optimizing for speed to reduce management overhead.  The overall
compiler pipeline is shown in Figure~\ref{fig:compilerDiagram} and
the rest of this subsection covers this optimization pipeline.

\subsubsection{CKKS to OpenFHE}\label{Sec.Comp.CtO}
\begin{figure}[!htbp]
\centering
\includegraphics[width=0.58\columnwidth]{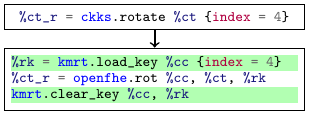}
\caption{Lowering from the target independent \texttt{ckks} dialect to \texttt{kmrt} and \texttt{openfhe} dialects. Highlighted lines show the added operations for loading and clearing rotation keys.  This pass is described in Section~\ref{Sec.Comp.CtO}.}
\label{fig:lowering_rot}
\end{figure}
%TODO abbreviations ending with dot \@
CKKS to OpenFHE is a builtin flow in HEIR which lowers \texttt{ckks} into \texttt{lwe} and \texttt{lwe} into \texttt{openfhe} dialects. We have extended this pass so that \texttt{kmrt} operations \texttt{kmrt.load\_key} and \texttt{kmrt.clear\_key} are 
inserted to manage key memories. 
As in Figure \ref{fig:lowering_rot}, this pass associates every rotation operation with a specific \texttt{kmrt.rot\_key} value. To reinforce this association, rotation operations in \texttt{openfhe} are extended to  require a rotation key typed value as an argument.

\textbf{Purpose} Every rotation operation gets its own locally live rotation key. This approach inverts the problem of creating liveness ranges to merging existing liveness ranges. This way, the compiler  focuses on reducing the number of key loadings by merging liveness ranges.

\subsubsection{Lowering Linear Transform}\label{Sec.Comp.LLT}
This pass legalizes the IR by rewriting  \texttt{linear\_transform} operations as an MLIR \texttt{affine} loop with its basic block consisting of primitive \texttt{ckks} operations such as \texttt{add}, \texttt{multiply} and \texttt{rotate}. The lowering uses the diagonal order matrix vector multiplication structure which is a commonly used matrix-vector multiplication algorithm in FHE~\cite{haleviFasterHomomorphicLinear2018}. This generates rotations with indices ranging from $1$ to vector size $N$.

\textbf{Purpose} Expand \texttt{linear\_transform} op into supported FHE operations.

\subsubsection{Baby-Step Giant-Step Decomposition}\label{Sec.Comp.BSGS}
This pass reduces the number of keys required from $N$ to $~\sqrt{N}$ by chaining rotations in BSGS structure.
% \subfile{code_rotation_decomposition}
The pass finds \texttt{affine} loop index ranges from $1$ to $N$ and tiles the loop in BSGS structure that results in chained rotations for a maximum of two. This results in two nested loops with the outer loop performing giant steps and inner loop performing the baby step rotation and the main multiply-add computation. Section \ref{Sec.Background.RC}, explains the details of BSGS. BSGS decomposition is preferred for matrix-vector multiplication patterns because of its key reduction benefits and low additional rotation cost. Even though KeyMemRT allows running at server side without BSGS, large number of keys causes excessive key generation times at client and high disk storage at server.

The nested loop structure presents an opportunity for key management as the inner loop with baby steps is known to use the same rotation keys at every full iteration while the outer loop uses each giant step key once. Inner loop keys from $1$ to $~\sqrt{N}$ are kept in memory in the loop execution while the outer loop loads and clears the key it needs.  This reduces memory pressure with minimal key management overhead. We use the BSGS Decomposition pass together with Merge Rotation Keys (\ref{Sec.Comp.MRK}) to detect key reuse across kernels and optimize the key liveness ranges.

\textbf{Purpose} Reduces number of keys needed for consecutive rotation indices by chaining rotations in BSGS structure.

\subsubsection{Merge Rotation Keys}\label{Sec.Comp.MRK}

\begin{figure}[!htbp]
\centering
\includegraphics[width=0.6\columnwidth]{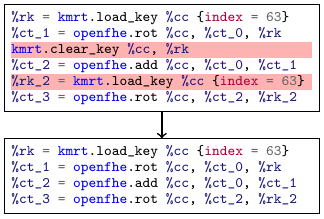}
\caption{This code shows two rotation operations with the same index and thus the same key. In the unoptimized snippet (top), the same key is loaded, cleared and right after one operation loaded again. This optimization looks out for opportunities where the same key is cleared and loaded in near distance and removes the pair.  This optimization is discussed further in Section~\ref{Sec.Comp.MRK}}
\label{fig:merge_clear_deser}
\end{figure}
This pass detects sequences of clearing and loading rotation key with the same index with dataflow analysis and removes them if they are close in distance.

Combination of CKKS to OpenFHE and Lower Linear Transform provide numerous liveness range merging opportunities. Keys that are frequently used explicitly by the program or the linear transform are detected and kept in memory longer. Figure \ref{fig:merge_clear_deser} shows a simple example of explicit key reuse that results in liveness merging. 

\textbf{Purpose} This way redundant loads are removed and latency is optimized. The pass is used as a key management canonicalizer for other passes that create redundant load and clear operations.

\subsubsection{Rotation Hoisting}\label{Sec.Comp.RH}
Rotation operation internally has a stage that is only input dependent that can be reused with 
different indices when hoisted~\cite{haleviFasterHomomorphicLinear2018}. This pass finds the patterns
of rotations applied to same input and transforms them into faster rotations supported by OpenFHE. The pass
accelerates rotations in a loop when their input is a common loop invariant 
with loop hoisting.

\textbf{Purpose} This pass speeds up programs by transforming patterns of rotations into faster alternatives.

% \clearpage
\subsubsection{Common Subexpression Elimination}\label{Sec.Comp.CSE}
Common Subexpression Elimination (CSE) is used to remove cleartext dead code resulting
from \texttt{kmrt} optimizations. It is used to simplify the IR and reduce code size.

\textbf{Purpose} Remove cleartext dead code.

\subsubsection{Bootstrap Removal}\label{Sec.Comp.BR}
This pass uses profiling information to get ciphertext level information 
from the program and removes unnecessary bootstrap operations. If a bootstrap operation
consumes more levels than it obtains, it is considered unnecessary. This pass detects 
changes in ciphertext levels to identify and remove these bootstraps. Since bootstrap is a 
time-heavy operation, it speeds up programs.

\textbf{Purpose} Identify and remove bootstraps that do not contribute to the program.

\subsubsection{Bootstrap Key Management}\label{Sec.Comp.BKM}
This pass adds \texttt{kmrt} memory management operations around bootstrap operations for its internal rotations.
In this pass, first the rotation indices that the bootstrap operation needs is inferred by checking its setup parameters. Then the indices found are inserted as key loading and clearing ops wrapping around bootstrap operations exposing bootstrap keys in the IR. 

The keys materialized with this pass allow the succeeding "Merge Rotation Keys" pass to perform holistic program-level key management by detecting key reuse among explicit rotations, linear transforms and bootstraps.

\textbf{Purpose} Materialize bootstrap keys and add key management to reduce average memory consumption. 

\subsubsection{Insert Clear Ops}\label{Sec.Comp.ICO}
The pass reduces the memory footprint of FHE program variables such as ciphertexts and plaintexts. For this, it performs a basic liveness analysis for all such variables and insert a clear op after their last use. 

\textbf{Purpose} Reduce memory impact of FHE ciphertext and plaintext variables.

\subsubsection{Hint Placement for Key Prefetching}\label{Sec.Comp.KP}

Key prefetching aims to overlap computations with key loadings by inserting hint operations for the runtime. KeyMemRT compiler inserts \texttt{prefetch\_key} hint operations before \linebreak \texttt{load\_key} operations to inform the runtime of upcoming loads. We prefer to delegate the prefetching timing to the runtime so the compiler simply provides the list of keys to be loaded as a sequence of \texttt{prefetch\_key} operations in the beginning of the program.

\textbf{Purpose} Placing hints to inform the runtime of the upcoming key loadings to enable asynchronous computation and key movement.

\section{KeyMemRT Runtime}

KeyMemRT Runtime connects the compiler analysis of key lifetimes with the FHE library and the running machine. The runtime provides several APIs that the compiler translates \texttt{kmrt} operations to. The runtime has access to the dynamic FHE library context and can change it on the fly. It also has access to the machine's active memory usage information and its capacity which are crucial information for portability of the optimizations.

\begin{figure}[!htbp]
\centering
\includegraphics[width=0.6\columnwidth]{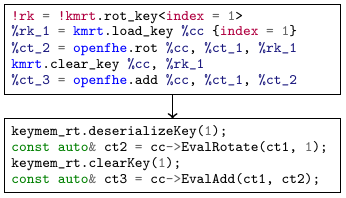}
\caption{An example of MLIR to C++ translation.  The \texttt{kmrt} operations become C++ calls in the runtime to manage rotation keys.}
\label{fig:mlir_translation}
\end{figure}

After KeyMemRT Compiler runs its optimizations on the \texttt{kmrt} and target specific \texttt{openfhe} MLIR dialects, they are translated to their C++ corresponding code. As in Figure \ref{fig:mlir_translation}, \texttt{openfhe} dialect is translated into OpenFHE library calls and \texttt{kmrt} operations are translated to calls to the KeyMemRT Runtime. Below its APIs, the runtime communicates with the OpenFHE library's context.

KeyMemRT Runtime runs alongside the FHE application and it can monitor system resources. The runtime has two modes:

\begin{description}
	\item[Low-Memory Mode]: The runtime executes directives in order and sequentially. This mode ensures minimal memory usage but adds key-management overhead time to the total execution time.
	\item[Balanced Mode]: This mode allows a balance between latency and memory usage by loading keys in parallel with the computation and ahead-of-time. By collaborating with the compiler's hint placement operations, the runtime starts to load keys before they are needed. The runtime makes sure a key is available when \texttt{load\_key} operation is encountered. The loading thread continually loads from an internal queue and takes into account the loaded number of keys. When the loaded keys are saturated up to the user given limit, the runtime waits for clear operations to open space for new keys.
\end{description}

\subsection{Runtime Memory Management}

\begin{figure}
	   \includegraphics[width=\columnwidth]{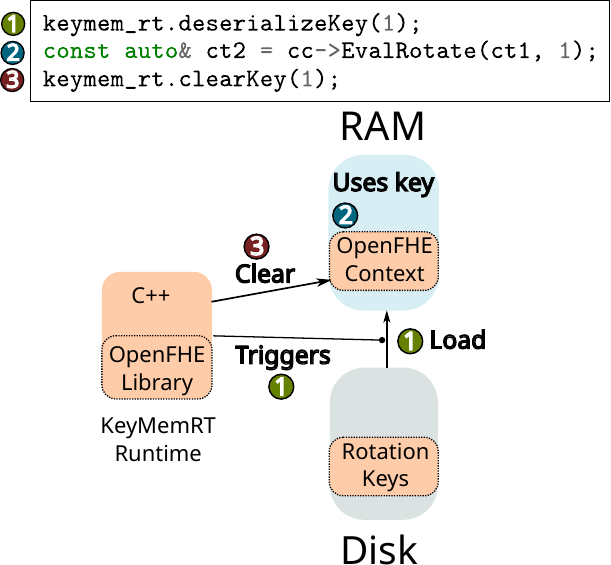}
	   \caption{KeyMemRT Runtime. The generated C++ code calls the runtime which loads keys and clears from memory when required.}
	   \label{fig:runtime_diagram}
\end{figure}

In a typical FHE client-server architecture, the rotation keys 
are transferred from client to the server with other necessary 
cryptographic data. In server execution, FHE programs expect 
 the rotation keys to be stored in disk. For this
architecture, FHE libraries provide APIs for clients to generate 
keys in memory and serialize (store) them to disk from memory; and for servers to
deserialize (load) keys from disk to memory.

KeyMemRT Runtime workflow is visualized in Figure \ref{fig:runtime_diagram}. The optimized server instead of directly
loading keys with the library APIs, instead defers rotation key loading 
and management to the KeyMemRT runtime. 
\includegraphics[height=2ex]{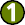} When the program hits a KeyMemRT call, the runtime gets triggered and depending on its mode and internal state, calls the library level key loading API.
The runtime triggers
the library to copy the needed specific key from disk into library context active in memory. 
\includegraphics[height=2ex]{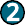} The library can use the loaded key in 
rotation operations that is supported by the key.  
\includegraphics[height=2ex]{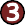} When there is no need to keep the
key in memory, the program informs the runtime to clear the key. The runtime 
selectively clears the key from the context's internal map in memory. Since the key
is not removed from disk but copied it can be loaded from there again so there 
is no need to store back.
 KeyMemRT runtime APIs provide the 
 backbone of the infrastructure that is targetable by its compiler.

\section{Results}
We evaluate the KeyMemRT pipeline on multiple ML models
exploring memory consumption and execution time (Section~\ref{sec:memory_time}).
We show a geomean memory reduction of \memImprAntace{} compared
to ANT-ACE
and a latency improvement of \timeImprFhelipe{} compared to Fhelipe.

\subsection{Setup}

We run the experiments in 
a server environment with Intel Xeon Gold 6154 CPU with 72 cores at 3GHz and with 512 GB of RAM\@.
Our OpenFHE fork is based on v1.2.3 with minor changes to better support serialization.
We offload rotation key handling to the KeyMemRT runtime and
rely on OpenFHE for the FHE computations and general functionality.

\textbf{Benchmarks}  We evaluate \name{} on multiple ML models ranging from small to large sizes
for FHE execution. We used the same FHE parameters across benchmarks 
with ring dimension of $2^{16}$ and multiplicative depth of 30. Due to these parameters, each key take up around 130 MB in size. Section \ref{Sec.Results.KeySize} shows how increases in ring dimension or multiplicative depth can increase key sizes significantly. Table \ref{tab:network_keys} shows how many keys each program uses, along with the total memory requirements from the keys alone.

\subsubsection{KeyMemRT Modes}
We evaluate two modes of KeyMemRT: \textbf{Low-Memory} and \textbf{Balanced}.
\textbf{Low-Memory} focuses on memory reduction, delaying key loads until they are
required to minimize the memory footprint at the cost of latency.
\textbf{Balanced} enables prefetching, which reduces the overheads
that the KeyMemRT runtime introduces, but increase the memory consumption slightly.
We compare these two modes in depth across the evaluation.

\subsubsection{Alternative Approaches}
We evaluate against two different state-of-the-art
FHE compilers, ANT-ACE and Fhelipe.  These are summarized
below:
\begin{description}[nolistsep,noitemsep]
	\item[ANT-ACE]
		ANT-ACE has a global approach to rotation keys,
		keeping all rotation keys for all indices of the program
		alive for the duration of the program.
	\item[Fhelipe]
		Fhelipe uses powers-of-two rotation chaining, which
		reduces the number of live-keys but increases execution time.
\end{description}
We evaluate against these approaches in the same server environment
with the same model and memory requirements.

\subsection{Memory Consumption and Execution Time}\label{sec:memory_time}

\begin{figure*}
	\includegraphics[width=\textwidth]{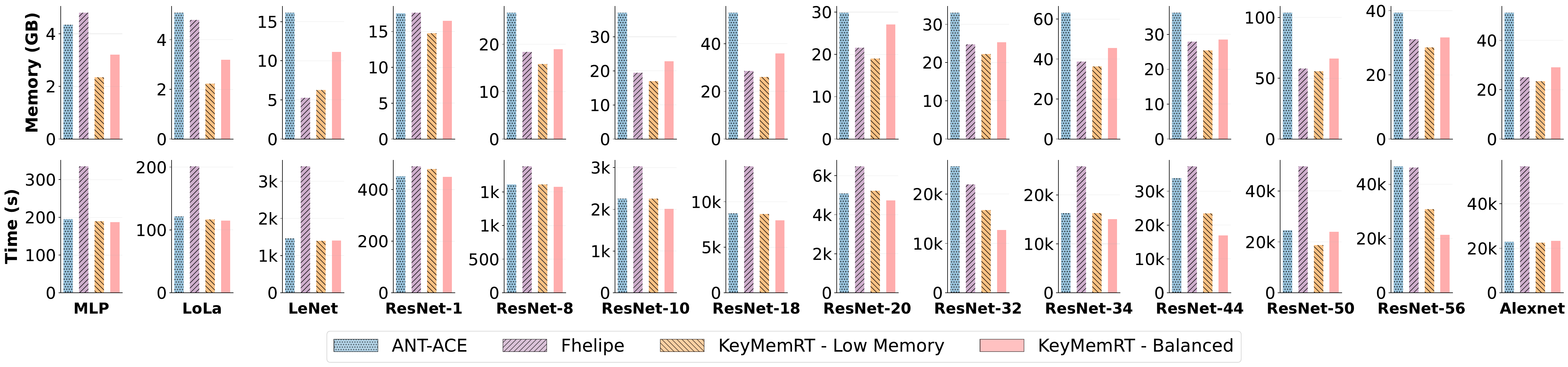}
   \caption{
	   Peak memory and execution time comparison in ML models. KeyMemRT reduces ANT-ACE's memory consumption by up to 2.6x (LeNet) with a geomean memory reduction of \memImprAntace{}. KeyMemRT speeds up over Fhelipe's execution by up to 2.5x (AlexNet) with a geomean speedup of \timeImprFhelipe{}.}
   
	\label{fig:time_mem_all_models}
\end{figure*}

  \begin{table}
  \centering
  \caption{Programs and the number of rotation keys each needs}
  \label{tab:network_keys}
  \begin{tabular}{|l|c|c|}
  \hline
  \textbf{Program} & \textbf{Number of} & \textbf{Memory (GB)} \\
   & \textbf{Keys} &  \\
  \hline
  MLP & 27 & 3.5 \\
  LoLa & 33 & 4.3 \\
  ResNet-1 & 75 & 9.8 \\
  LeNet & 117  & 15.2 \\
  ResNet-8,20,32,44,56 & 139 & 18.1 \\
  ResNet-10 & 213 & 27.7 \\
  ResNet-18,34 & 267 & 34.7 \\
  AlexNet & 285 & 37.1 \\
  ResNet-50 & 442 & 57.5 \\
  \hline
  \end{tabular}
  \end{table}

Figure~\ref{fig:time_mem_all_models} shows how different key management strategies of FHE compilers affect
memory requirements and runtimes of individual programs.  We see geomean reduction in memory
footprint of \memImprAntace{} compared to ANT-ACE due to our key management strategy.  We see
geomean execution time improvement of
\timeImprFhelipe{} over Fhelipe's memory reduction strategy because KeyMemRT avoids duplicating
computation to achieve low-memory consumption.

ANT-ACE shows the highest peak memory utilization, while KeyMemRT Low Memory mode shows the lowest in most benchmarks. Fhelipe is the second lowest in large benchmarks such as the ResNet variants or AlexNet.

\subsubsection{Small Programs}\label{sec:small_progs}
 MLP, LoLa and LeNet are depth-wise small benchmarks that do not consume excessive multiplicative depth and can run without bootstraps. These benchmarks do not include bootstrap keys and can have lower memory footprint.

\textbf{Memory}:
 ANT-ACE shows memory consumption proportional to the number of keys of each model while Fhelipe is consuming a fixed amount of memory around 5~GB which are mostly due to the keys supporting powers-of-two rotations. KeyMemRT Low Memory mode significantly improves memory of ANT-ACE ranging between 1.85-2.59x. KeyMemRT Balanced mode follows the behaviour of Low Memory mode with additional memory overhead due to prefetching. The improvements are due to KeyMemRT keeping only a subset of rotation keys of ANT-ACE active at any given time.

 Since LeNet requires many more keys than MLP or LoLa, ANT-ACE struggles to scale memory and uses it excessively as it allows projecting number of keys to the memory consumption. KeyMemRT optimizations have a live set of keys that is much less than ANT-ACE that translates to a significant memory reduction of 2.59x. Although Fhelipe uses the least memory, it pays for this with very high
 compute time.

\textbf{Time}:
ANT-ACE and both KeyMemRT modes show similar latency performance with KeyMemRT showing slight speedups ranging between 1.03-1.05x. However, Fhelipe is significantly slower in all benchmarks by a large margin. Fhelipe is slower by 1.7x in MLP and LoLa, and 2.4x slower in LeNet. In LeNet, having more number of keys increases the likelihood of having larger integers to express as sums of powers-of-two in longer sequences. Fhelipe's memory reduction strategy severely harms its  latency performance. On the other hand, KeyMemRT demonstrates that it is possible to reduce memory to a level similar or better than Fhelipe without regressing on performance. KeyMemRT does not add the overhead of Fhelipe as it still allows the program to execute every rotation in a single step similar to ANT-ACE as it keeps the exact keys needed instead of a predetermined sequence like powers-of-two.

\subsubsection{Large Programs}\label{sec:large_progs}
ResNet variants and AlexNet are large FHE programs as they require bootstrap operations to continue their execution by refreshing their depths. This adds the bootstrap keys to all of the compiled benchmarks increasing memory footprint. 

\textbf{Memory}:
ANT-ACE shows the highest memory consumption that struggles to scale in large programs such as in ResNet-50 surpassing 100~GB of memory usage. Key intensive programs such as AlexNet also show excessive memory consumption with ANT-ACE. KeyMemRT reduces memory consumption of ANT-ACE ranging between 1.19-2.19x. Fhelipe closely follows KeyMemRT Low Memory mode however the lack of bootstrap key management limits the potential memory lowering it can do.

ResNet-8,20,32,44,56 models have a similar ML architecture but have varying repetitiveness. For this reason, they have the same number of keys and the absolute memory improvement from keys with KeyMemRT are thus close to each other at around 10~GBs.

ResNet-10,18,34,50 and AlexNet have very high number of keys and ANT-ACE reflects the rise as an increase in memory consumption. KeyMemRT provides significant improvements to these models as it can remove most of the key memory overhead of ANT-ACE.

\textbf{Time}:
ResNet-1,10,18,20,34,50 and AlexNet benchmarks follow the pattern of having similar latencies for ANT-ACE and KeyMemRT modes, with Balanced mode having the best speed in most cases. Fhelipe slows down significantly by around 2.5x in ResNet-50 and AlexNet with more varied indices stressing the rotation chain lengths.

Having similar ML architectures with different depths, ResNet-8,20,32,44,56 models show increasing speedups with KeyMemRT in order. The combined improvements in memory reduction and KeyMemRT's Balanced mode prefetching demonstrate that it is possible to increase performance in some memory-bounded scenarios with programs that better inform the underlying system of the upcoming data movement. Balanced mode demonstrates large performance improvement compared to ANT-ACE by 1.99-2.18x

%NOTE We have a compiler bug at ResNet-50 Low Memory Mode so the data is incomplete, I ignore it in text but we should rerun with either manual fixing to the output c++ or actually starting from the compiler.
Models with very high number of keys, ResNet-10,18,34,50 and AlexNet show Fhelipe struggling with high numbers of keys. As the integers for indices get larger, the chain of rotations that Fhelipe uses to express a single rotation can get longer translating into regression in performance. ANT-ACE and KeyMemRT Low Memory mode shows similar latency performance while KeyMemRT Balanced mode having a slight speedup.

\label{sec:memory_results}
%\subsubsection{Comparison to Existing Frameworks}

%\subsubsection{A Deep-Dive into \name{}'s Memory Requirements}
% KeyMemRT's memory optimizations are program dependent and can 
% have different characteristics through time. Figure \ref{fig:resnet_memory_over_time} shows
% memory utilization of ResNet-8 in time.
% We can compare KeyMemRT Low Memory mode
% to Fhelipe and see it has lower memory utilisation most of the time and has lower execution time.
% Another comparison can be drawn from ANT-ACE and KeyMemRT Balanced mode.
% While ANT-ACE takes shorter time, the memory utilisation is considerably higher than KeyMemRT Balanced mode.
%
% We can also see the key changes throughout the program, including the large
% increase in memory utilization at the start of the program and the peaks
% within the program for bootstrapping, which requires a large number of keys.

% \begin{figure*}[t]
% 	\includegraphics[width=\textwidth]{results/alexnet_memory_over_time.pdf}
% 	\caption{Alexnet execution memory profile over time with a warm-start.  We can see the effects
% 	of active key management, with lower startup loading,
% 	and peaks for bootstraps which require a large number of keys.  }
% 	\label{fig:alexnet_memory_over_time}
% \end{figure*}

\subsection{Key Sizes}\label{Sec.Results.KeySize}

\begin{figure}
	\includegraphics[width=\columnwidth]{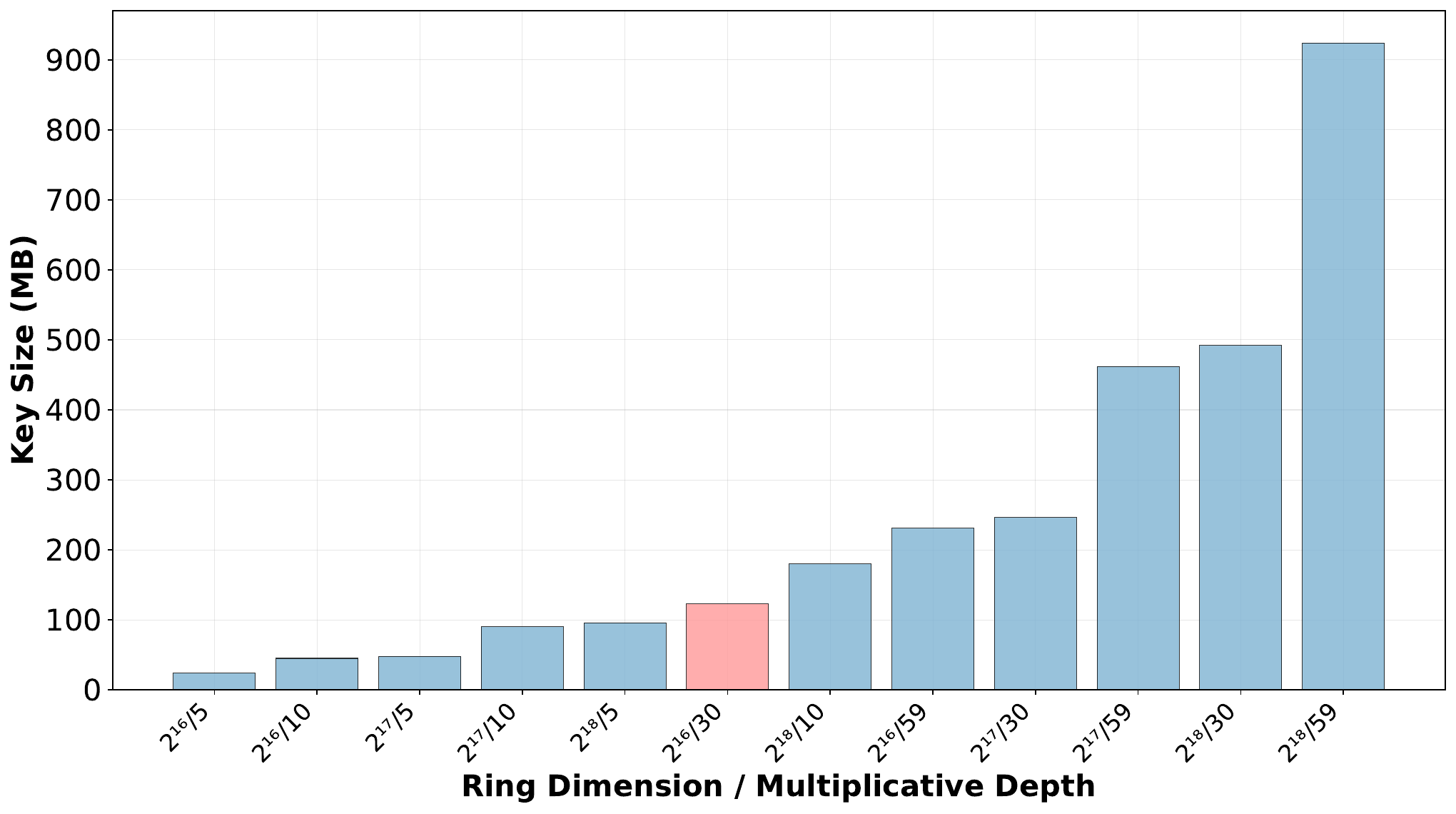}
	\caption{Size of a single rotation key depends on FHE parameters such as Ring Dimension and Multiplicative Depth. The highlighted column shows the configuration that this work uses.}
	\label{fig:key_sizes}
\end{figure}

Rotation Key sizes can change drastically depending on the FHE parameters selected. Figure \ref{fig:key_sizes} shows measurements from the serialized keys generated from OpenFHE. Doubling ring dimension with the same depth, doubles the key size. Multiplicative depth is also highly correlated with the key size as higher depth will cause a larger rotation key size. This work uses ring dimension of $2^{16}$ and depth of $30$ however it is common to use higher depths depending on the algorithm. It is possible to use higher ring dimension for increased cryptographic security or to have more slots in an FHE vector.

While this work provides memory improvements in the given configuration, it is applicable to all sizes of rotation keys.

\section{Related Work}
A number of FHE frameworks exist to optimize compute
resource, but research on reducing memory consumption
has been sparse.  Here, we cover this related work.

\subsection{FHE Compilers}

A number of FHE compilers generate FHE operations from higher-level
descriptions~\cite{viandSoKFullyHomomorphic2021a}.
%There has been a number of compilers dedicated to compiling FHE programs from 
%higher level descriptions to FHE operations \cite{viandSoKFullyHomomorphic2021a}.
FHE compilers lower the barrier for programming FHE by automating cryptographic
decisions. FHE compilers also aim to increase performance of FHE applications as they 
can better exploit FHE constructs.

\textbf{Reducing Number of Rotations}. Rotom~\cite{chenBridgingUsabilityPerformance2025},
Fhelipe~\cite{krastevTensorCompilerAutomatic2024}, Orion ~\cite{ebelOrionFullyHomomorphic2025} 
, ANT-ACE ~\cite{liANTACEFHECompiler2025}, CHET ~\cite{dathathriCHETOptimizingCompiler2019}, EVA ~\cite{dathathriEVAEncryptedVector2020} and HEIR ~\cite{aliHEIRUniversalCompiler2025} are FHE compilers that specifically focus on mapping
Machine Learning models to high performance FHE programs. These compilers optimize the program 
by reducing the number of rotation operations.
While reducing number
of rotations improves latency,
the memory impact is dependent on the number of indices removed which these
compilers do not target.

HECO~\cite{viandHECOFullyHomomorphic2023a}, Coyote~\cite{malikCoyoteCompilerVectorizing2023} and Porcupine~\cite{cowanPorcupineSynthesizingCompiler2021b} are general purpose FHE compilers that generate FHE kernels with fewer rotations, again saving time, but not reducing memory consumption.

\textbf{Rotation Chaining}. Rotom, Orion and ANT-ACE make use of Baby-Step Giant-Step decomposition
that reduces the required number of rotation keys. Fhelipe dynamically evaluates a rotation index
as a series of binary powers. This means Fhelipe generated programs to run more slowly.

\textbf{Manual Key Management}. Orion uses hand-written coarse-grained
key loading and clearing to reduce its memory footprint.

\textbf{No Key Management}. ANT-ACE, HEIR, CHET,
HECO, Coyote and Porcupine do not have key management which creates memory 
scalability problems.

\textbf{Other Compilers}. HEaaN.MLIR~\cite{parkHEaaNMLIROptimizingCompiler2023}, HECATE~\cite{leeHECATEPerformanceAwareScale2022} are FHE compilers that optimize different parts of FHE programs than rotations. CHESS~\cite{shokriCHESSCompilingHomomorphic2025}, PEGASUS~\cite{luPEGASUSBridgingPolynomial2021} and HEIR~\cite{bianHEIRUnifiedRepresentation2024} are compilers focused on hybrid use of different types of FHE schemes.  Concrete~\cite{Concrete}, Parasol~\cite{ParasolCompilerPushing} and PyTFHE~\cite{maPyTFHEEndtoEndCompilation2023} are compilers targeting bitwise FHE schemes which have different operation sets defined. 

\subsection{FHE Libraries}

FHE libraries provide the implementation for FHE operations including rotation
and storage for keys. Some libraries address the memory issue caused by rotation
keys. 

\textbf{Rotation Chaining}. SEAL~\cite{chenSimpleEncryptedArithmetic2017a} library implement rotation 
chaining for rotation indices that have no keys generated. Its chaining logic is
based on a special representation called Non-Adjacent Form that extends over binary
representation. It uses negative powers to have lower number of rotations but keeps additional 
rotation keys for these indices globally. HELib~\cite{haleviFasterHomomorphicLinear2018} provides support for BSGS decomposition. 

Although rotation keys cause large memory issues, libraries are not able to provide definitive
solutions since they are at a lower level than programs. Rotation indices depend on the FHE program
 which determine the necessary key set. However, FHE libraries and frameworks have provided optimizations
 for inner level phases of rotation operations and rotation keys.

 \textbf{Key-Switching Optimization.} Level-aware key-switching \cite{hwangOptimizingHEOperations2023}, CiFlow~\cite{nedaCiFlowDataflowAnalysis2024}, Lattigo library~\cite{bossuatEfficientBootstrappingApproximate2021} are examples of 
 optimizing the key-switching phase of rotation operations.

 \textbf{Transciphering.} Transciphering focuses on client-side key generation performance and communication cost of rotation keys\cite{leeRotationKeyReduction2023, cheonLightweightCKKSClient, hwangOptimizingHEOperations2023}. Hierarchical Rotation Key System \cite{leeRotationKeyReduction2023}, allows the client to send a small number of rotation keys with minimal communication overhead. Hierarchical rotation keys do not reduce the memory footprint at the server side but instead shift the key generation load from client to server. 

 \textbf{Key Size Reduction.} HELib~\cite{haleviDesignImplementationHElib2020} reduces key sizes by not storing pseudo-random part of rotation keys and instead generating it on-the-fly. 

These optimizations allow more keys in the same space but do not solve the memory-scalability problem of rotation keys. They are orthogonal to this work and are out of scope.

\subsection{FHE Accelerators}

Due to constrained memories, research on accelerators often discusses the memory scalability issue of rotation keys.

\textbf{Rotation Chaining} ARK~\cite{kimARKFullyHomomorphic2022} uses Min-KS,  a Horner-like rule~\cite{haleviFasterHomomorphicLinear2018} for minimising the number of rotation keys. This is an extreme example of rotation chaining where a single rotation key is stored and rotations are recursively calculated. Min-KS works on consecutive ranges similar to BSGS.

\textbf{Key Size Reduction.} GPU specific optimizations \cite{jung100xFasterBootstrapping2021, kimCheddarSwiftFully2024} involve selecting FHE parameters that favor smaller sizes of keys but increases operation level computation complexity.

\section{Conclusion}
This paper introduces \name{}, a compiler
and runtime framework that reduces the memory footprint
of large-scale FHE programs by \memImprAntace{}.
\name{} is the first FHE compiler to implement and automate
fine-grained key management strategies and bootstrap
key management.

The memory optimizations enable more efficient
use of compute and network resources, unlocking better
throughput for large FHE programs. We evaluate ML models and demonstrate a memory reduction of \memImprAntace{} and a speedup of \timeImprAntace{} over ANT-ACE; and memory reduction of \memImprFhelipe{} and speedup of \timeImprFhelipe{} over memory-optimized Fhelipe.

\bibliographystyle{abbrv}
\bibliography{FHE}

\begin{thebibliography}{10}

\bibitem{ParasolCompilerPushing}
Parasol {{Compiler}}: {{Pushing}} the {{Boundaries}} of {{FHE Program
  Efficiency}}.

\bibitem{aliHEIRUniversalCompiler2025}
A.~Ali, J.~Choi, B.~Gipson, S.~Gorantala, J.~Kun, W.~Legiest, L.~Lim, A.~Viand,
  M.~Z. Demissie, and H.~Zheng.
\newblock {{HEIR}}: {{A Universal Compiler}} for {{Homomorphic Encryption}},
  Aug. 2025.

\bibitem{badawiOpenFHEOpenSourceFully2022}
A.~A. Badawi, A.~Alexandru, J.~Bates, F.~Bergamaschi, D.~B. Cousins,
  S.~Erabelli, N.~Genise, S.~Halevi, H.~Hunt, A.~Kim, Y.~Lee, Z.~Liu,
  D.~Micciancio, C.~Pascoe, Y.~Polyakov, I.~Quah, S.~R.V, K.~Rohloff,
  J.~Saylor, D.~Suponitsky, M.~Triplett, V.~Vaikuntanathan, and V.~Zucca.
\newblock {{OpenFHE}}: {{Open-Source Fully Homomorphic Encryption Library}},
  2022.

\bibitem{bianHEIRUnifiedRepresentation2024}
S.~Bian, Z.~Zhao, Z.~Zhang, R.~Mao, K.~Suenaga, Y.~Jin, Z.~Guan, and J.~Liu.
\newblock {{HEIR}}: {{A Unified Representation}} for {{Cross-Scheme
  Compilation}} of {{Fully Homomorphic Computation}}.
\newblock In {\em Proceedings 2024 {{Network}} and {{Distributed System
  Security Symposium}}}, San Diego, CA, USA, 2024. Internet Society.

\bibitem{bossuatEfficientBootstrappingApproximate2021}
J.-P. Bossuat, C.~Mouchet, J.~{Troncoso-Pastoriza}, and J.-P. Hubaux.
\newblock Efficient {{Bootstrapping}} for {{Approximate Homomorphic
  Encryption}} with {{Non-sparse Keys}}.
\newblock In A.~Canteaut and F.-X. Standaert, editors, {\em Advances in
  {{Cryptology}} -- {{EUROCRYPT}} 2021}, pages 587--617, Cham, 2021. Springer
  International Publishing.

\bibitem{brakerskiFullyHomomorphicEncryption2012}
Z.~Brakerski.
\newblock Fully {{Homomorphic Encryption}} without {{Modulus Switching}} from
  {{Classical GapSVP}}.
\newblock In R.~{Safavi-Naini} and R.~Canetti, editors, {\em Advances in
  {{Cryptology}} -- {{CRYPTO}} 2012}, volume 7417, pages 868--886. Springer
  Berlin Heidelberg, Berlin, Heidelberg, 2012.

\bibitem{brakerskiFullyHomomorphicEncryption}
Z.~Brakerski, C.~Gentry, and V.~Vaikuntanathan.
\newblock Fully {{Homomorphic Encryption}} without {{Bootstrapping}}.

\bibitem{chenBridgingUsabilityPerformance2025}
E.~Chen, F.~Brown, and W.~Zheng.
\newblock Bridging {{Usability}} and {{Performance}}: {{A Tensor Compiler}} for
  {{Autovectorizing Homomorphic Encryption}}, 2025.

\bibitem{chenSimpleEncryptedArithmetic2017}
H.~Chen, K.~Laine, and R.~Player.
\newblock Simple {{Encrypted Arithmetic Library}} - {{SEAL}} v2.1.
\newblock In M.~Brenner, K.~Rohloff, J.~Bonneau, A.~Miller, P.~Y. Ryan,
  V.~Teague, A.~Bracciali, M.~Sala, F.~Pintore, and M.~Jakobsson, editors, {\em
  Financial {{Cryptography}} and {{Data Security}}}, volume 10323, pages 3--18.
  Springer International Publishing, Cham, 2017.

\bibitem{chenSimpleEncryptedArithmetic2017a}
H.~Chen, K.~Laine, and R.~Player.
\newblock Simple {{Encrypted Arithmetic Library}} - {{SEAL}} v2.1, 2017.

\bibitem{cheonLightweightCKKSClient}
J.~H. Cheon, M.~Kang, and J.~H. Park.
\newblock Towards {{Lightweight CKKS}}: {{On Client Cost Efficiency}}.

\bibitem{cheonHomomorphicEncryptionArithmetic2017}
J.~H. Cheon, A.~Kim, M.~Kim, and Y.~Song.
\newblock Homomorphic {{Encryption}} for {{Arithmetic}} of {{Approximate
  Numbers}}.
\newblock In T.~Takagi and T.~Peyrin, editors, {\em Advances in {{Cryptology}}
  -- {{ASIACRYPT}} 2017}, volume 10624, pages 409--437. Springer International
  Publishing, Cham, 2017.

\bibitem{cowanPorcupineSynthesizingCompiler2021b}
M.~Cowan, D.~Dangwal, A.~Alaghi, C.~Trippel, V.~T. Lee, and B.~Reagen.
\newblock Porcupine: A synthesizing compiler for vectorized homomorphic
  encryption.
\newblock In {\em Proceedings of the 42nd {{ACM SIGPLAN International
  Conference}} on {{Programming Language Design}} and {{Implementation}}},
  {{PLDI}} 2021, pages 375--389, New York, NY, USA, June 2021. Association for
  Computing Machinery.

\bibitem{dathathriEVAEncryptedVector2020}
R.~Dathathri, B.~Kostova, O.~Saarikivi, W.~Dai, K.~Laine, and M.~Musuvathi.
\newblock {{EVA}}: An encrypted vector arithmetic language and compiler for
  efficient homomorphic computation.
\newblock In {\em Proceedings of the 41st {{ACM SIGPLAN Conference}} on
  {{Programming Language Design}} and {{Implementation}}}, {{PLDI}} 2020, pages
  546--561, New York, NY, USA, June 2020. Association for Computing Machinery.

\bibitem{dathathriCHETOptimizingCompiler2019}
R.~Dathathri, O.~Saarikivi, H.~Chen, K.~Laine, K.~Lauter, S.~Maleki,
  M.~Musuvathi, and T.~Mytkowicz.
\newblock {{CHET}}: An optimizing compiler for fully-homomorphic neural-network
  inferencing.
\newblock In {\em Proceedings of the 40th {{ACM SIGPLAN Conference}} on
  {{Programming Language Design}} and {{Implementation}}}, {{PLDI}} 2019, pages
  142--156, New York, NY, USA, June 2019. Association for Computing Machinery.

\bibitem{ebelOrionFullyHomomorphic2025}
A.~Ebel, K.~Garimella, and B.~Reagen.
\newblock Orion: {{A Fully Homomorphic Encryption Framework}} for {{Deep
  Learning}}.
\newblock In {\em Proceedings of the 30th {{ACM International Conference}} on
  {{Architectural Support}} for {{Programming Languages}} and {{Operating
  Systems}}, {{Volume}} 2}, pages 734--749, Rotterdam Netherlands, Mar. 2025.
  ACM.

\bibitem{fanSomewhatPracticalFully2012b}
J.~Fan and F.~Vercauteren.
\newblock Somewhat {{Practical Fully Homomorphic Encryption}}, 2012.

\bibitem{haleviFasterHomomorphicLinear2018}
S.~Halevi and V.~Shoup.
\newblock Faster {{Homomorphic Linear Transformations}} in {{HElib}}.
\newblock In H.~Shacham and A.~Boldyreva, editors, {\em Advances in
  {{Cryptology}} -- {{CRYPTO}} 2018}, volume 10991, pages 93--120. Springer
  International Publishing, Cham, 2018.

\bibitem{haleviDesignImplementationHElib2020}
S.~Halevi and V.~Shoup.
\newblock Design and implementation of {{HElib}}: A homomorphic encryption
  library, 2020.

\bibitem{hwangOptimizingHEOperations2023}
I.~Hwang, J.~Seo, and Y.~Song.
\newblock Optimizing {{HE}} operations via {{Level-aware Key-switching
  Framework}}.
\newblock In {\em Proceedings of the 11th {{Workshop}} on {{Encrypted
  Computing}} \& {{Applied Homomorphic Cryptography}}}, pages 59--67,
  Copenhagen Denmark, Nov. 2023. ACM.

\bibitem{jung100xFasterBootstrapping2021}
W.~Jung, S.~Kim, J.~H. Ahn, J.~H. Cheon, and Y.~Lee.
\newblock Over 100x {{Faster Bootstrapping}} in {{Fully Homomorphic
  Encryption}} through {{Memory-centric Optimization}} with {{GPUs}}.
\newblock {\em IACR Transactions on Cryptographic Hardware and Embedded
  Systems}, pages 114--148, Aug. 2021.

\bibitem{kimCheddarSwiftFully2024}
J.~Kim, W.~Choi, and J.~H. Ahn.
\newblock Cheddar: {{A Swift Fully Homomorphic Encryption Library}} for {{CUDA
  GPUs}}, July 2024.

\bibitem{kimARKFullyHomomorphic2022}
J.~Kim, G.~Lee, S.~Kim, G.~Sohn, M.~Rhu, J.~Kim, and J.~H. Ahn.
\newblock {{ARK}}: {{Fully Homomorphic Encryption Accelerator}} with {{Runtime
  Data Generation}} and {{Inter-Operation Key Reuse}}.
\newblock In {\em 2022 55th {{IEEE}}/{{ACM International Symposium}} on
  {{Microarchitecture}} ({{MICRO}})}, pages 1237--1254, Oct. 2022.

\bibitem{kimOptimizedRankSort2025}
S.~Kim, E.~{\"U}nay, A.~{Yilmazer-Metin}, and H.~T. Lee.
\newblock Optimized {{Rank Sort}} for {{Encrypted Real Numbers}}, 2025.

\bibitem{krastevTensorCompilerAutomatic2024}
A.~Krastev, N.~Samardzic, S.~Langowski, S.~Devadas, and D.~Sanchez.
\newblock A {{Tensor Compiler}} with {{Automatic Data Packing}} for {{Simple}}
  and {{Efficient Fully Homomorphic Encryption}}.
\newblock {\em Proc. ACM Program. Lang.}, 8(PLDI):152:126--152:150, June 2024.

\bibitem{lattnerMLIRScalingCompiler2021}
C.~Lattner, M.~Amini, U.~Bondhugula, A.~Cohen, A.~Davis, J.~Pienaar, R.~Riddle,
  T.~Shpeisman, N.~Vasilache, and O.~Zinenko.
\newblock {{MLIR}}: Scaling compiler infrastructure for domain specific
  computation.
\newblock In {\em Proceedings of the 2021 {{IEEE}}/{{ACM International
  Symposium}} on {{Code Generation}} and {{Optimization}}}, {{CGO}} '21, pages
  2--14, Virtual Event, Republic of Korea, Sept. 2021. IEEE Press.

\bibitem{leeRotationKeyReduction2023a}
J.-W. Lee, E.~Lee, Y.-S. Kim, and J.-S. No.
\newblock Rotation {{Key Reduction}} for~{{Client-Server Systems}} of~{{Deep
  Neural Network}} on~{{Fully Homomorphic Encryption}}.
\newblock In J.~Guo and R.~Steinfeld, editors, {\em Advances in {{Cryptology}}
  -- {{ASIACRYPT}} 2023}, pages 36--68, Singapore, 2023. Springer Nature.

\bibitem{leeRotationKeyReduction2023}
J.-W. Lee, E.~Lee, Y.-S. Kim, and J.-S. No.
\newblock Rotation {{Key Reduction}} for~{{Client-Server Systems}} of~{{Deep
  Neural Network}} on~{{Fully Homomorphic Encryption}}.
\newblock In J.~Guo and R.~Steinfeld, editors, {\em Advances in {{Cryptology}}
  -- {{ASIACRYPT}} 2023}, pages 36--68, Singapore, 2023. Springer Nature.

\bibitem{leeHECATEPerformanceAwareScale2022}
Y.~Lee, S.~Heo, S.~Cheon, S.~Jeong, C.~Kim, E.~Kim, D.~Lee, and H.~Kim.
\newblock {{HECATE}}: {{Performance-Aware Scale Optimization}} for
  {{Homomorphic Encryption Compiler}}.
\newblock In {\em 2022 {{IEEE}}/{{ACM International Symposium}} on {{Code
  Generation}} and {{Optimization}} ({{CGO}})}, pages 193--204, Seoul, Korea,
  Republic of, Apr. 2022. IEEE.

\bibitem{liANTACEFHECompiler2025}
L.~Li, J.~Lai, P.~Yuan, T.~Sui, Y.~Liu, Q.~Zhu, X.~Zhang, L.~Xiao, W.~Chen, and
  J.~Xue.
\newblock {{ANT-ACE}}: {{An FHE Compiler Framework}} for {{Automating Neural
  Network Inference}}.
\newblock In {\em Proceedings of the 23rd {{ACM}}/{{IEEE International
  Symposium}} on {{Code Generation}} and {{Optimization}}}, {{CGO}} '25, pages
  193--208, New York, NY, USA, Mar. 2025. Association for Computing Machinery.

\bibitem{luPEGASUSBridgingPolynomial2021}
W.-j. Lu, Z.~Huang, C.~Hong, Y.~Ma, and H.~Qu.
\newblock {{PEGASUS}}: {{Bridging Polynomial}} and {{Non-polynomial
  Evaluations}} in {{Homomorphic Encryption}}.
\newblock In {\em 2021 {{IEEE Symposium}} on {{Security}} and {{Privacy}}
  ({{SP}})}, pages 1057--1073, May 2021.

\bibitem{maPyTFHEEndtoEndCompilation2023}
J.~Ma, C.~Xu, and L.~W. Wills.
\newblock {{PyTFHE}}: {{An End-to-End Compilation}} and {{Execution Framework}}
  for {{Fully Homomorphic Encryption Applications}}.
\newblock In {\em 2023 {{IEEE International Symposium}} on {{Performance
  Analysis}} of {{Systems}} and {{Software}} ({{ISPASS}})}, pages 24--34, Apr.
  2023.

\bibitem{malikCoyoteCompilerVectorizing2023}
R.~Malik, K.~Sheth, and M.~Kulkarni.
\newblock Coyote: {{A Compiler}} for {{Vectorizing Encrypted Arithmetic
  Circuits}}.
\newblock In {\em Proceedings of the 28th {{ACM International Conference}} on
  {{Architectural Support}} for {{Programming Languages}} and {{Operating
  Systems}}, {{Volume}} 3}, {{ASPLOS}} 2023, pages 118--133, New York, NY, USA,
  Mar. 2023. Association for Computing Machinery.

\bibitem{marcollaSurveyFullyHomomorphic2022}
C.~Marcolla, V.~Sucasas, M.~Manzano, R.~Bassoli, F.~H.~P. Fitzek, and N.~Aaraj.
\newblock Survey on {{Fully Homomorphic Encryption}}, {{Theory}}, and
  {{Applications}}, 2022.

\bibitem{mazzoneEfficientRankingOrder2025}
F.~Mazzone, M.~Everts, F.~Hahn, and A.~Peter.
\newblock Efficient {{Ranking}}, {{Order Statistics}}, and {{Sorting}} under
  {{CKKS}}.
\newblock In {\em 34th {{USENIX Security Symposium}} ({{USENIX Security}} 25)},
  pages 8541--8558, 2025.

\bibitem{mouchetLattigoMultipartyHomomorphic2020}
C.~V. Mouchet, J.-P. Bossuat, J.~R. {Troncoso-Pastoriza}, and J.-P. Hubaux.
\newblock Lattigo: {{A}} multiparty homomorphic encryption library in {{Go}}.
\newblock In {\em Proceedings of the 8th {{Workshop}} on {{Encrypted
  Computing}} and {{Applied Homomorphic Cryptography}}}, pages 64--70, 2020.

\bibitem{nedaCiFlowDataflowAnalysis2024}
N.~Neda, A.~Ebel, B.~Reynwar, and B.~Reagen.
\newblock {{CiFlow}}: {{Dataflow Analysis}} and {{Optimization}} of {{Key
  Switching}} for {{Homomorphic Encryption}}.
\newblock In {\em 2024 {{IEEE International Symposium}} on {{Performance
  Analysis}} of {{Systems}} and {{Software}} ({{ISPASS}})}, pages 61--72, May
  2024.

\bibitem{parkHEaaNMLIROptimizingCompiler2023}
S.~Park, W.~Song, S.~Nam, H.~Kim, J.~Shin, and J.~Lee.
\newblock {{HEaaN}}.{{MLIR}}: {{An Optimizing Compiler}} for {{Fast Ring-Based
  Homomorphic Encryption}}.
\newblock {\em Proc. ACM Program. Lang.}, 7(PLDI):114:196--114:220, June 2023.

\bibitem{paszkePyTorchImperativeStyle2019}
A.~Paszke, S.~Gross, F.~Massa, A.~Lerer, J.~Bradbury, G.~Chanan, T.~Killeen,
  Z.~Lin, N.~Gimelshein, L.~Antiga, A.~Desmaison, A.~K{\"o}pf, E.~Yang,
  Z.~DeVito, M.~Raison, A.~Tejani, S.~Chilamkurthy, B.~Steiner, L.~Fang,
  J.~Bai, and S.~Chintala.
\newblock {{PyTorch}}: An imperative style, high-performance deep learning
  library.
\newblock In {\em Proceedings of the 33rd {{International Conference}} on
  {{Neural Information Processing Systems}}}, number 721, pages 8026--8037.
  Curran Associates Inc., Red Hook, NY, USA, Dec. 2019.

\bibitem{shokriCHESSCompilingHomomorphic2025}
R.~Shokri and N.~G. Tsoutsos.
\newblock {{CHESS}}: {{Compiling Homomorphic Encryption}} with {{Scheme
  Switching}}.
\newblock In {\em 2025 {{IEEE International Symposium}} on {{Hardware Oriented
  Security}} and {{Trust}} ({{HOST}})}, pages 324--334, May 2025.

\bibitem{viandHECOFullyHomomorphic2023a}
A.~Viand, P.~Jattke, M.~Haller, and A.~Hithnawi.
\newblock {{HECO}}: {{Fully Homomorphic Encryption Compiler}}.
\newblock In {\em 32nd {{USENIX Security Symposium}} ({{USENIX Security}} 23)},
  pages 4715--4732, 2023.

\bibitem{viandSoKFullyHomomorphic2021a}
A.~Viand, P.~Jattke, and A.~Hithnawi.
\newblock {{SoK}}: {{Fully Homomorphic Encryption Compilers}}.
\newblock In {\em 2021 {{IEEE Symposium}} on {{Security}} and {{Privacy}}
  ({{SP}})}, pages 1092--1108, May 2021.

\bibitem{Concrete}
{Zama}.
\newblock Concrete: {{TFHE Compiler}} that converts python programs into
  {{FHE}} equivalent, 2022.

\end{thebibliography}

%%%%%%%%%%%%%%%%%%%%%%%%%%%%%%%%%%%%%%%%%%%%%%%%%%%%%%%%%%%%%%%%%%%%%%%%%%%%%%%%
\end{document}